\documentclass[preprint,5p]{elsarticle}

\usepackage{lineno,hyperref}
\usepackage{graphicx}
\usepackage{subfig}
\usepackage{amsmath,amsthm}
\usepackage{epsfig}
\usepackage{bm}
\usepackage{balance}
\usepackage{mathtools}
\usepackage{epstopdf}

\usepackage{amssymb}

\modulolinenumbers[5]
\journal{Journal of Theoretical Biology}

\biboptions{authoryear}

\begin{document}

\begin{frontmatter}

\title{Coupling geometry on binary bipartite networks:
                     	         hypotheses testing on pattern geometry and nestedness}

\author[mymainaddress]{ Jiahui Guan }
\ead{jiaguan@ucdavis.edu}

\author[mymainaddress]{Hsieh Fushing\corref{mycorrespondingauthor}}
\cortext[mycorrespondingauthor]{Corresponding author}
\ead{fhsieh@ucdavis.edu}

\address[mymainaddress]{Department of Statistics, University of California, Davis}

\begin{abstract}
Upon a matrix representation of a binary bipartite network, via the permutation invariance, a coupling geometry is computed to approximate the minimum energy macrostate of a network's system. Such a macrostate is supposed to constitute the intrinsic structures of the system, so that the coupling geometry should be taken as information contents, or even the nonparametric minimum sufficient statistics of the network data. Then pertinent null and alternative hypotheses, such as nestedness, are to be formulated according to the macrostate. That is, any efficient testing statistic needs to be a function of this coupling geometry. These conceptual architectures and mechanisms are by and large still missing in community ecology literature, and rendered misconceptions prevalent in this research area. Here the algorithmically computed coupling geometry is shown consisting of deterministic multiscale block patterns, which are framed by two marginal ultrametric trees on row and column axes, and stochastic uniform randomness within each block found on the finest scale. Functionally a series of increasingly larger ensembles of matrix mimicries is derived by conforming to the multiscale block configurations. Here matrix mimicking is meant to be subject to constraints of row and column sums sequences. Based on such a series of ensembles, a profile of distributions becomes a natural device for checking the validity of testing statistics or structural indexes. An energy based index is used for testing whether network data indeed contains structural geometry.  A new version block-based nestedness index is also proposed. Its validity is checked and compared with the existing ones. A computing paradigm, called Data Mechanics, and its application on one real data network are illustrated throughout the developments and discussions in this paper.

\end{abstract}


\end{frontmatter}

\section{Introduction}

 Ever since the “assembly rules” proposed in \cite{diamond1975assembly} and \cite{case1983pattern}, the presence-absence matrix has been the fundamental data type in Community Ecology. A presence-absence matrix is also called co-occurrence matrix. In fact, from data structural perspective, as being permutation invariant on both axes of matrix, this kind of data type should be precisely termed binary bipartite network. Such network data is now a major data type for understanding mutualistic system interactions in wide ranges of ecological studies (\cite{bascompte2003nested}). A ecological mutualistic system concerns mutually beneficial interactions between a collection of animal species and another collection of plant species. One typical example is the flowering plants and their insect pollinators. The binary bipartite network records presence-absence of a target interaction upon each animal- vs-plant entry. In other words, a binary bipartite network is used to approximate an ecological system from mutualistic perspective. In contrast, the directed binary network is used to approximate an antagonistic system, such as a food web. It should be noted that these two binary networks have rather distinct structures and information contents. They can't be mixed. The directed network is not considered here.

Given its fundamental role, the controversy centering on the binary bipartite network has never faded away for the past four decades. One key reason underlying this controversy, to our opinion, is that the intrinsic mathematical architectures and proper physical mechanisms underlying such binary bipartite networks are by and large still missing in current ecological literature. The consequences of this missing include: 1) observed pattern within a binary bipartite network has never been identified analytically (\cite{connor1978species}); 2) network or matrix based structural hypotheses are not precisely formulated (\cite{gotelli2000null, ulrich2007null} ); 3) validity of testing statistics is not properly checked (\cite{diamond1975assembly,stone1990checkerboard}); 4) and the computations of p-value for statistical inferences are apparently incorrect (\cite{diamond1982examination, gotelli2000null, ulrich2007null}). All these consequences are caused in part by lacking of knowledge of information contents contained within an observed binary bipartite network, and in part lacking proper algorithms for mimicking and generating matrices with distinct structural information. Above all challenging tasks of defining an effective testing statistics on matrix or network data have not been systematically resolved. Only heuristic and parsimonious solutions are suggested so far in literature.

All these aforementioned issues are systematically discussed, developed and resolved in this paper. Throughout this paper a binary bipartite network and its approximating mutualistic system are the primary concerns. Therefore the network's rectangle matrix representation has one axis for a collection of animals of interest and the other for a collection of plants under study. We first discuss computational developments for what visible geometric patterns are indeed embedded within the matrix, and then discuss whether such embedded geometric structures are coherent with the idea of nestedness. The first part of discussion resolves the issues arising from co-occurrence matrices observed in biogeographic systems.

By making use of the fact that a binary bipartite network is permutation invariant with respect to nodes on both axes, a new computing paradigm, called Data Mechanics, is applied to extract a combination of a deterministic multiscale structures and a stochastic uniformity from such data (\cite{fushing2014data}). The coupling of deterministic and stochastic structures is termed a coupling geometry. This resultant coupling geometry is taken as the computable information contents of the network data because it is very close to the minimum energy macrostates of the target system. From statistical physics perspective, all microstates are to conform to such macrostates.  Such a conformation implies a principle of how to mimic an observed network data (\cite{fushing2014data}). Specifically the deterministic multiscale structures are the visible patterns contained in the data, which are what have been missing in  (\cite{connor1978species}), while the uniformity enables us to mimic and to generate various ensembles of matrices with different geometric pattern information.

Another major concept proposed in this paper is that the conceptual nestedness on a data matrix has to be adapted upon the computed deterministic multiscale structures. This adaptation is meant to build the least nestedness-bearing construct containing observed deterministic multiscale structures. As such testing the hypothesis of whether a data matrix embedding with geometry of nestedness is to evaluate the degree of structural differences between this nestedness-bearing construct and the original coupling geometry's deterministic structures. Based on this concept, we propose a block based nestedness index and compare it with three existing popular indexes. Among these three indexes, one is originally proposed in \cite{patterson1986nested} and the other two are the improved versions (\cite{almeida2008consistent, atmar1993measure}). Ironically we found that these two improved versions are indeed improper. Throughout this paper we use the well-studied Mammal data in (\cite{patterson1986nested}) for illustrating and expositional purposes.

\section{Method}
\subsection{From intuitive grouping ideas to coupling geometry }
\begin{figure*}[t]
\centering
\subfloat[]{\includegraphics[trim=0cm 0cm 0cm 0cm, clip, width=3.2in]{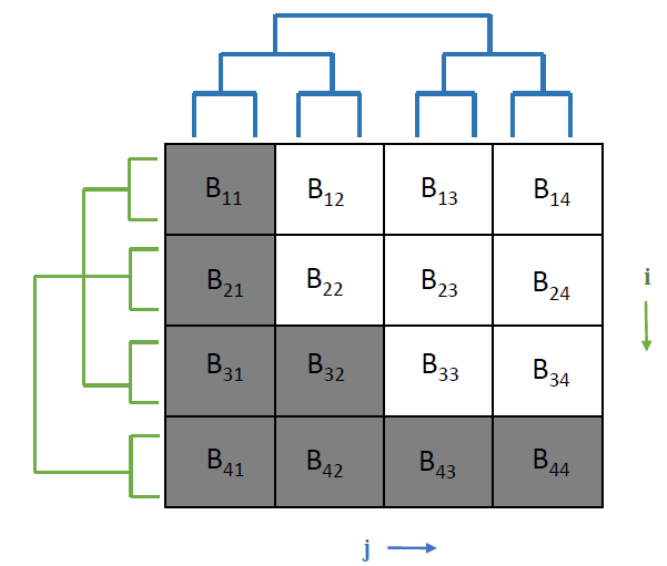}}
\subfloat[]{\includegraphics[trim=0cm 0cm 0cm 0cm, clip, width=3.2in]{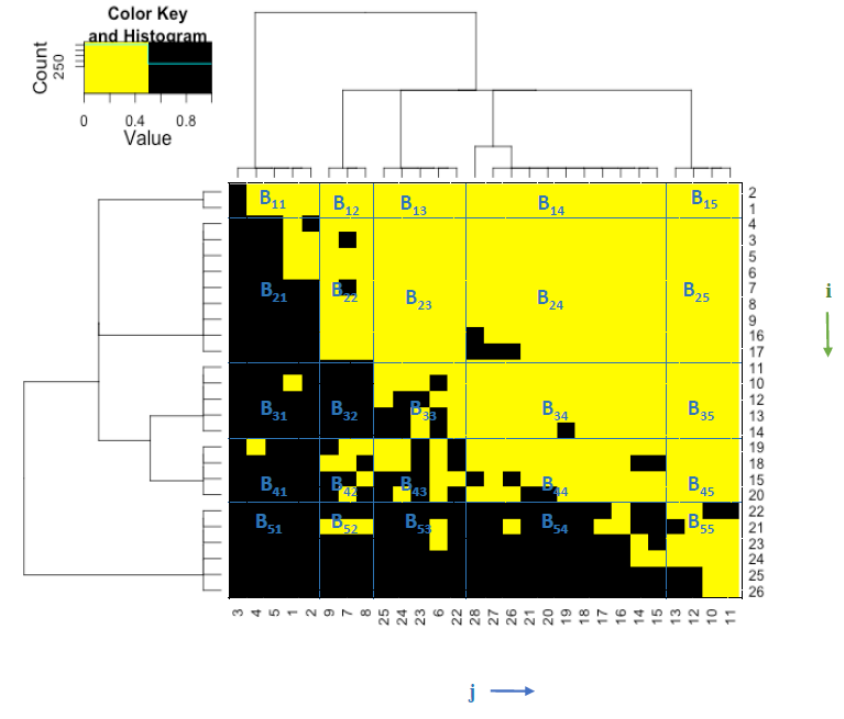}}

\caption{(a) illustration of theoretical multi-scale block structures and Coupling Geometry of a perfect nested bipartite network. Interactions are marked in grey while non-interaction are in white. (b) the multi-scale block structures and Coupling Geometry of Mammalian dataset that contains 26 mammalian species in 28 mountain ranges [\cite{patterson1986nested}]. Interactions are in black and non-interaction are in yellow.}
\label{fig:Figure1}
\end{figure*}

 Within an ecological system, all intrinsic patterns of the mutualistic interactions between a collection of animal species and another collection of plant species, beyond individual-to-individual level, are the supposed information contents to be contained within the observed binary bipartite network. Such pattern information intuitively would be jointly expressed through clusters of similar animals coupling with clusters of similar plants in a fashion of block-wise uniformity. That is, on multiple global levels, dissimilar clusters on one axis would reveal contrasting configurations of clusters on the other axis. As such, scientists can visualize why different clusters of animal are characterized distinctively with respect to differences among clusters of plant. That is, the information contents within a binary bipartite network data are multiscale and visible, more importantly they are computable.

Hierarchical clustering can somehow capture and visualize block-wise clustering of a matrix, but it tends to produce clusters with imbalanced sizes and each block lacks uniformity (\cite{johnson1967hierarchical}). Recently such multiscale information patterns are computed through a new computing paradigm, called Data Mechanics, developed in \cite{fushing2014data}. Computationally, Data Mechanics indeed attempts to solve an optimal permutation problem of achieving the minimum total variation among all possible matrix- representations of the observed bipartite network. Here the total variation is defined with respect to a choice of neighborhood system, such as the set of immediate neighbors on the rectangle matrix lattice. A version of total variation with detailed formula is given in Supplementary Section A. This discrete combinatorial optimization is operated based on the permutation invariance of a bipartite network with respect to its nodes of animals and plants. The complexity of this problem surely depends on the exponentially growing factorials of sizes of the animal and plan collections. Though the concept of pattern information contents contained on a binary bipartite network is intuitive, the computing for the multiscale structures can be a rather complex problem. Data Mechanics is designed to provide optimal or nearly optimal solutions to this computational problem.

The algorithm for computing ultrametric trees, a key part of Data Mechanics, is called Data Cloud Geometry (DCG). Developed in \cite{fushing2010time}, DCG is aimed to construct ultrametric trees via on multiscale clustering, which has been widely used in many fields (\cite{balasubramaniam2018social,gong2017sinogram}). Another important aspect of Data Mechanics is the iterative computation of ultrametric trees. Iterative algorithm has been proven that it can reduce systemetical errors and improve overal performance on many domains (\cite{gong2016assessment,gong2017iterative,li1998iterative}). With the iterative computing of DCG on row and column axes, Data Mechanics converts unstructured binary biparite networks into multiscale block patterns framed by two ultrametric trees iteratively built upon the two axes, respectively.

The stochastic structures are found within each block formed by a core cluster on row axis and one core cluster on the column axis. Such two-dimensional uniform randomness is subject to row and column sums sequences of the involving block. Here core clusters of an Ultrametric tree are identified on its bottom tree level. That is, the finest scale structure of a coupling geometry is referring to block patterns formed via core clusters, while the coarsest structure referring to the one framed by one cluster containing all animals and one cluster containing all plants. The scales between these two extremes are specified by tree levels between the top and the bottom one.

Thus, by designing all resultant optimal and nearly optimal solutions, we illustrate multiscale block patterns through the coupled framework of two ultrametric trees built on animal and plants axes in Fig.\ref{fig:Figure1} (a) and (b).

It is clear that the data-driven deterministic multiscale block structures brought out by multiple tree levels of two Ultrametric trees frame and summarize the interacting relational patterns between animals and plants. The coupling relation of these two trees in fact is derived iteratively and alternatively by applying a computing algorithm, called Data Cloud Geometry, which serves as the key device of the Data Mechanics. The iterative procedure is designed to update a distance measure used in the previous iteration by taking the currently computed tree structural information into considerations, while the procedure of alternating between animal and plant axes is designed to build the dependence or coupling of the two trees.

\begin{figure}[t]
\centering
\subfloat{\includegraphics[trim=0cm 0.5cm 0.4cm 0cm, clip, width=3.4in]{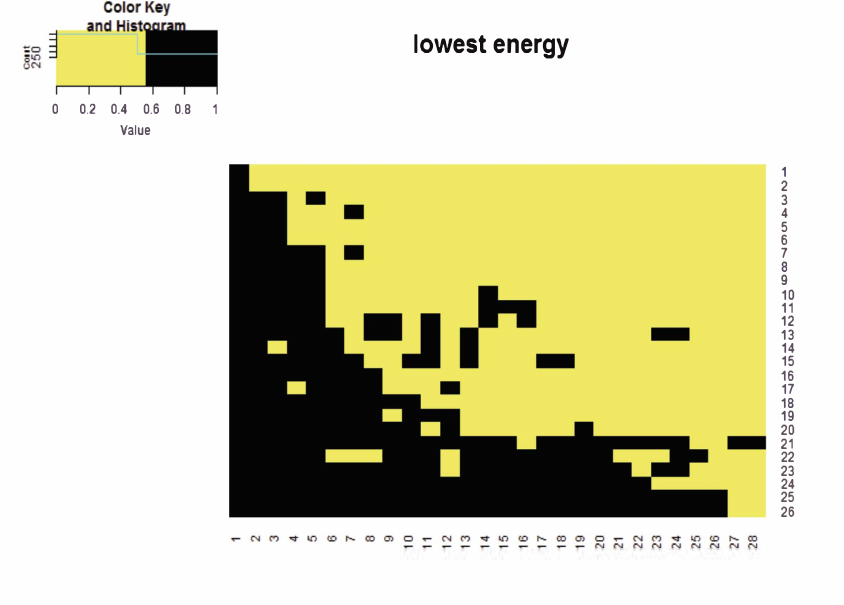}}
\caption{The heatmap of a binary matrix for the Mammalian dataset. This binary matrix has the lowest energy given that the row and column sums are fixed.}
\label{fig:Figure2}
\end{figure}

It is noted that throughout this paper a computed coupling geometry (with energy -2184), not the actually lowest energy matrix configurations, is employed as the foundation of all developments.  The reason for the parsimonious approach is purely for computational effectiveness. For instance, one lowest energy matrices (with energy -2204) of the Mammal data are found in Fig.\ref{fig:Figure2}.

It is typically that the computed coupling geometry is pretty close to the solution of lowest energy one. And it is the right starting point of searching endeavors of finding the optimal solution. However it should be noted that it usually takes a huge amount of computing efforts in order to achieve the optimal goal.

Beside the deterministic multiscale block structures, Data Mechanics computations also bring out block-wise stochastic randomness. This stochastic component is specifically seen as the uniformity within each block found on the finest scale.

\subsection{ From coupling geometry to block-based testing statistics for structural hypothesis}
Here we construct a reasonable and effective testing statistic in regarding to nestedness as a hypothesized geometric structure upon an observed binary matrix, or a binary bipartite network. Due to the fact that the coupling geometry is very close to the minimum energy macrostate of the system approximated by the data network, it is necessary to treat such a coupling geometry as the minimum sufficient statistic. It is a fundamental principle in statistical thinking that an efficient testing statistic should be based on the computed coupling geometry as the data's minimum sufficient statistic. Therefore the most relevant geometric structure of nestedness must be its least version that contains the coupling geometry. So theoretically finding the least containment is an optimization problem.

Let  $N_G$ denote the least nestedness geometric structure defined on the same matrix lattice as that of the originally observed data matrix. However there is no need to explicitly compute it because of the multiscale block patterns of the computed coupling geometry. Thus, we need only to evaluate the functional characteristics of $N_G$ in terms of all involving blocks, which are found on the finest scale upon a coupling geometry, as shown in Fig. \ref{fig:Figure1} and \ref{fig:Figure2}. That is, $\lambda (B_{ij}^{(N_G)})$ , the intensity of 1's in block $B_{ij}$ ,  has to satisfy the following two properties to be in accord with nestedness:

\begin{itemize}
\item[1)]	1st  order property:  $\{\lambda (B_{ij}^{(N_G)})\}$ is decreasing with respect to  given all $j$'s, and at the same time, increasing with respect to given all $i$'s.
\item[2)]	2nd order property: $\{\nabla \lambda (B_{ij}^{(N_G)}|C)\}$, 2nd order differences on  $i$th row:
\[\nabla \lambda (B_{ij}^{(N_G)}|C)=\lambda (B_{ij+1}^{(N_G)})-2\lambda (B_{ij}^{(N_G)})+\lambda (B_{ij-1}^{(N_G)})\]
has at most one sign-change from positive $(+)$ to negative $(-)$, that is, being concave down-ward to concave-upward; while $\{\nabla \lambda (B_{ij}^{(N_G)}|R)\}$, 2nd order differences on $j$th column:
\[\nabla \lambda (B_{ij}^{(N_G)}|R)=\lambda (B_{i+1j}^{(N_G)})-2\lambda (B_{ij}^{(N_G)})+\lambda (B_{i-1j}^{(N_G)})\]
has at most one sign-change from negative $(-)$ to positive $(+)$, that is, being concave up-ward to concave-downward.

\end{itemize}

Another important property of the second order is that sequences of  $\{\text{sign}(\nabla \lambda (B_{ij}^{(N_G)}|C))\}$ and
$\{\text{sign}(\nabla \lambda (B_{ij}^{(N_G)}|R))\}$ contain the corresponding sequences of signs pertaining to the coupling geometry, denoted as $C_G$. On the Mammal data, the $5\times 5$  matrix  $[\lambda(B_{ik}^{(N_G)})]$ of block-wise intensities of the coupling geometry is calculated as:
\begin{equation*}
\lambda(B_{ik}^{(C_G)}) =
  \begin{bmatrix}
  0.20 & 0.00 &  0.00 & 0.00 &0.00 \\
  0.84 & 0.07 &  0.00 & 0.04 &0.00\\
  0.96 & 1.00 &  0.36 & 0.02 &0.00 \\
  0.95 & 0.50 &  0.50 & 0.14 &0.00 \\
  1.00 & 0.83 &  0.93 & 0.89 &0.29 \\
  \end{bmatrix}
\end{equation*}
For $j=2,3,4; i= 1,2..5,$
\begin{equation*}
\text{sign}(\nabla \lambda (B_{ij}^{(N_G)}|C)) =
  \begin{bmatrix}
  +& + & + \\
  +& + & + \\
   - & + & + \\
  - & - & + \\
  - & - &  - \\
  \end{bmatrix}
\end{equation*}

For $i=2,3,4; j= 1,2..5,$
\begin{equation*}
\text{sign}(\nabla \lambda (B_{ij}^{(N_G)}|R)) =
  \begin{bmatrix}
  -& + & + &- &-\\
  +& + & + &+ & -\\
  +& + & + &+ & -\\
  \end{bmatrix}
\end{equation*}

Based on the above block-based nestedness perspective, the following nestedness-index for a simulated matrix, denoted by $S$ , is proposed:

\begin{small}

\begin{equation}
\begin{split}
&N_{CG}= \sum_i r_i \{ \sum_j [\sum_{k \neq j} (\lambda (B_{ij}^{(S)}-\lambda (B_{ik}^{(S)}))(j-k)]\}\\
& +\sum_jc_j \{\sum_i[ \sum_{k\neq i} (\lambda (B_{ij}^{(S)}-\lambda (B_{kj}^{(S)})
)(k-i)]\}\\
&- \sum_ir_i \{\sum_j(I-i+1)(j)\sum_{j>k>1}\{\nabla \lambda (B_{jk}^{(S)})\} \text{sign} (\nabla\lambda (B_{ik}^{(N_G)})|C)
\}\\
&- \sum_jc_j \{\sum_i(I-i+1)(j)\sum_{I>h>1}\{\nabla \lambda (B_{hj}^{(S)})\} \text{sign} (\nabla\lambda (B_{hj}^{(N_G)})|R)
\}\\
\end{split}
\end{equation}
\end{small}

The first two terms on the right hand sides of index $N_{CG}$ are “costs” against the linear ordering along the column-index on every row, and along the row-index on every column. The product terms are designed to be negative in values if the linear ordering holds, and positive if the linear ordering fails. So the larger $N_{CG}$  value is, the farther away from the nestedness. The 3rd and 4th terms are counting the coherence of 2nd order differences with that of $N_G$. Positive and larger values indicated incoherence or violations of nestedness.

\section{Result}
\subsection{From coupling geometry to matrix mimicking}
The primary use of a computed coupling geometry from a binary network is to make possible for generating a series of ensembles of matrix or network mimicries bearing with decreasing degrees of geometric structural information from the finest to the coarsest scales. The matrix ensemble pertaining to the finest scale of structural information is generated by patching up all simulated blocks, which are marked by core clusters of Ultrametric trees of animal and plants, subject to block-version row and column sums sequences. This is an ensemble that conforms to the minimum energy macrostate of the ecological system from statistical physics perspective. While the ensemble pertaining to the coarsest scale of structures is simply referring to the collection of matrices satisfying the constraints of row and column sums sequences of the observed entire matrix as a block.
\begin{figure}[t]
\centering
\subfloat{\includegraphics[trim=0cm 0.3cm 0cm 2cm, clip, width=3.6in]{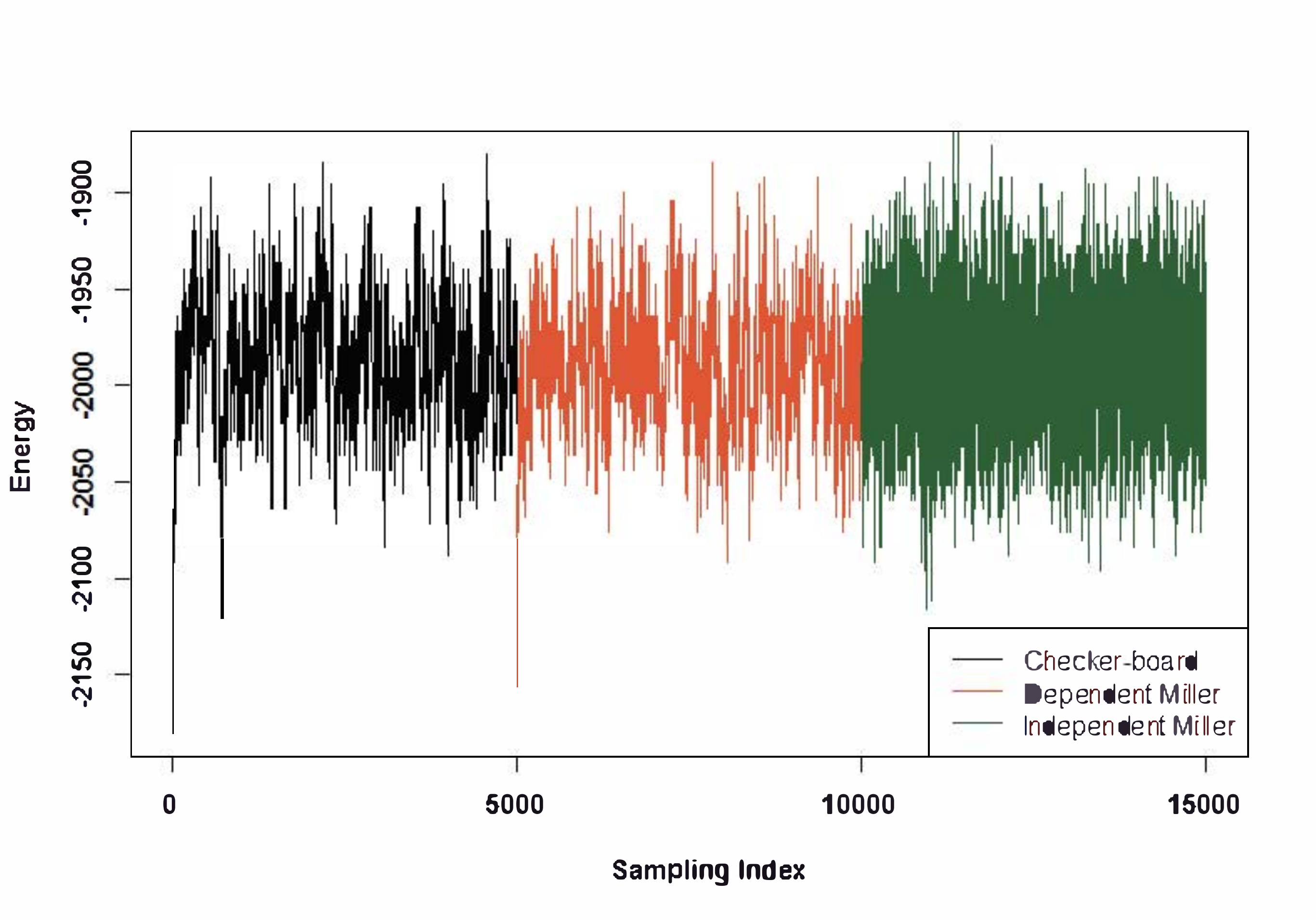}}
\caption{ Energy of total 15,000 sampling matrices with fixed row and column sums. Three different colors represent the sampling methods used: 2x2 checker-board, dependent Miller’s algorithm and independent Miller’s algorithm. }
\label{fig:Figure3}
\end{figure}
The generative algorithm employed here is the one proposed and used in \cite{miller2013exact}. A brief illustrating example and summary of this algorithm are given in the Supplementary Section B. But it is worth noting that this algorithm is effective for small sizes of binary data matrix, such as the Mammal data. It breaks down even on the $50\times 50$ matrix. The key factor affecting the performance of the algorithm is the matrix's sparsity of 1's.

In ecological literature, $2\times 2$ checker-board switching and its improved version, curveball algorithm (\cite{strona2014fast}), are also popularly used to generate binary matrices with constraints of row and column sums sequences. Basically the $2\times 2$ checker-board switching and its variants are searching for new solutions by going away from an existing one. In contrast, Miller and Harrison’s algorithm intrinsically simultaneously solving the linear equations imposed by the constraints of row and column sums sequences. Thus these two matrix generating algorithms are rather distinct in nature.  And both types of algorithms suffer distinct drawbacks to be applicable widely.

The drawbacks of $2\times 2$ checker-board switching and its variants are: first, they generate dependent matrices depending on the initial matrix; secondly, their energy spreads are relatively too narrow, indicating that they have preference for previously sampled matrix configurations. One evident view of such drawbacks is revealed in Fig.\ref{fig:Figure3}. Further our computer experiments show that the generating processes have rather short recurrent time cycles, that is, repeated matrices being generated too often.  This phenomenon indicates that the generated trajectory might have been confined within a small region.

Here we tentatively propose a practical way of resolving the issue of large data matrix that currently limiting Miller and Harrison’s algorithm.  By incorporating with a randomized divide-and-conquer sampling scheme on the observed data matrix, the whole matrix is divided into blocks, on which Miller and Harrison’s algorithm becomes applicable. This sampling scheme can be made to accommodate heterogeneity brought out the coupling geometry on both axes.

\subsection{From matrix ensembles to energy profile}
The entropies of this series of ensembles are defined as the logarithm of their sizes. For the Mammal data in Fig. \ref{fig:Figure1} (b), the serial sizes of ensembles are computed via an algorithm from \cite{miller2013exact} as follows: the size of the finest scale ($E_{5\times 5}$ version) ensemble is $1.3 \times 10^8$, the $E_{4\times 2}$version is $4.47 \times 10^{16}$, the $E_{2\times 2}$ version is $1.45 \times 10^{29}$ and the $E_{1\times 1}$ version (the coarsest scale one) is $2.7\times 10^{39}$. Such quantities of ensemble size or entropy bring out the quantitative sense of structural differences among multiscale block geometries embedded within the originally observed network.

\begin{figure*}[t]
\centering
\subfloat[]{\includegraphics[trim=0cm 0cm 0cm 1cm, clip, width=3.5in]{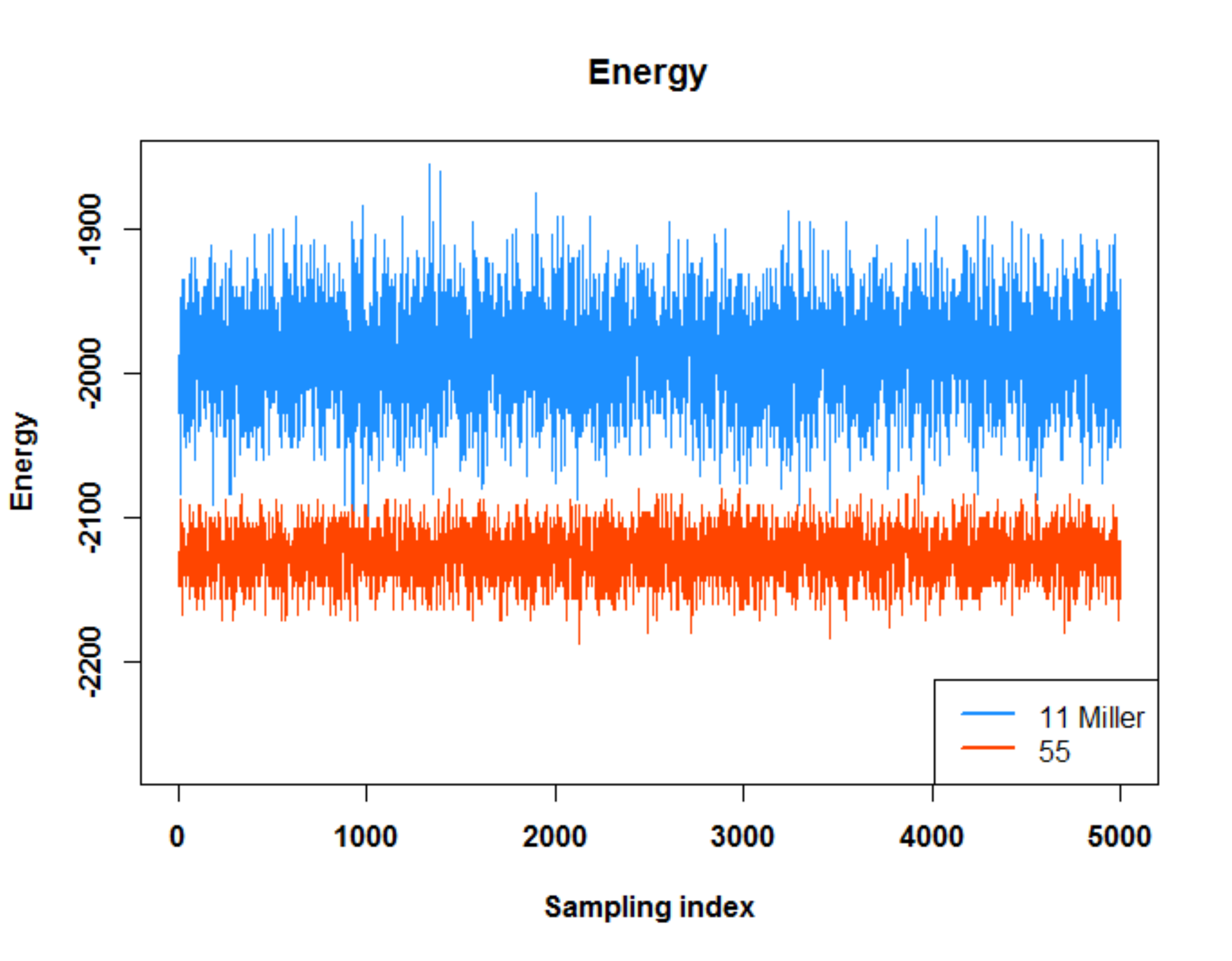}}
\subfloat[]{\includegraphics[trim=0cm 0cm 0cm 1cm, clip, width=3.4in]{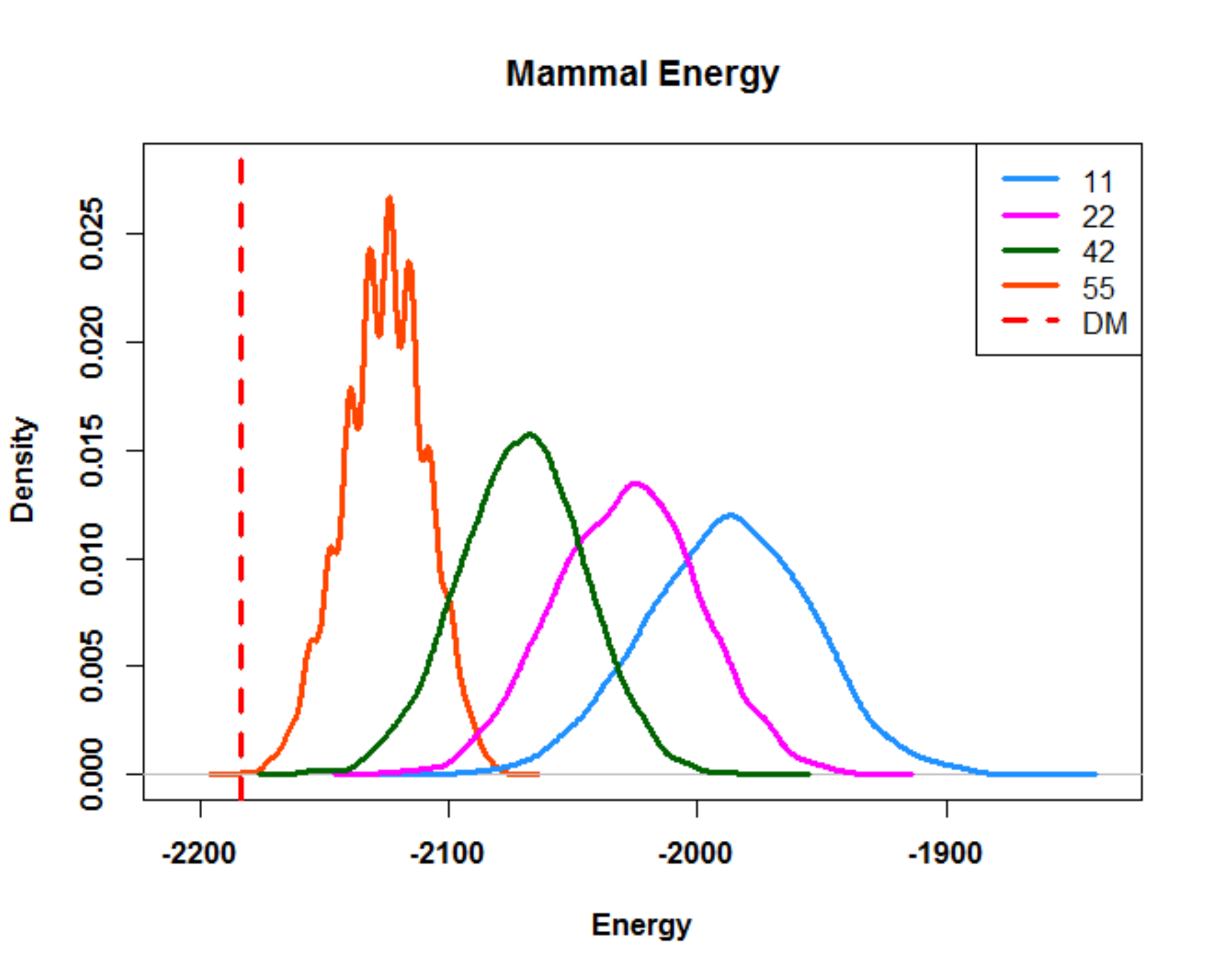}}

\caption{(a) Two energy trajectories from two scales of matrix strucutres. The light blue ones are sampled with fixed row and column sums of overall matrix, however, the orange-red ones are sampled with fixed row and column sums upon each $5\times 5$ blocks. (b) four energy distribution curves fit four different scales of block structures. }
\label{fig:Figure4}
\end{figure*}

Another important aspect of such differences is revealed via a profile of energy distributions, as shown in Fig. \ref{fig:Figure4} on the Mammal data.  It is evident that the two versions, $E_{5\times 5}$ and $E_{1\times 1}$, of ensembles are very different in a sense that a randomly chosen matrix from $E_{1\times 1}$ ensemble would appear very different from any one from$E_{5\times 5}$ensemble. The shifting-to-right pattern of the energy distribution profile strongly implies that computable and then observable block patterns contained in the coupling geometry is persistently eroding. The nearly complete separation of the two energy distributions based ensembles $E_{5\times 5}$ and $E_{1\times 1}$, respectively, indicates that the coupling geometry is not likely resulting from a random sampling.  In this fashion, the hypothesis of co-occurrence patterns is tested if the original binary bipartite network is represented by a presence-absence data matrix. (See the Supplementary Section A for comparisons of energy index with other indexes of co-occurrence.)

\subsection{	From coupling geometry to block-based testing statistics for structural hypothesis}
\begin{figure*}[t]
\centering
\subfloat[]{\includegraphics[trim=0cm 0.3cm 0cm 1.1cm, clip, width=3.5in]{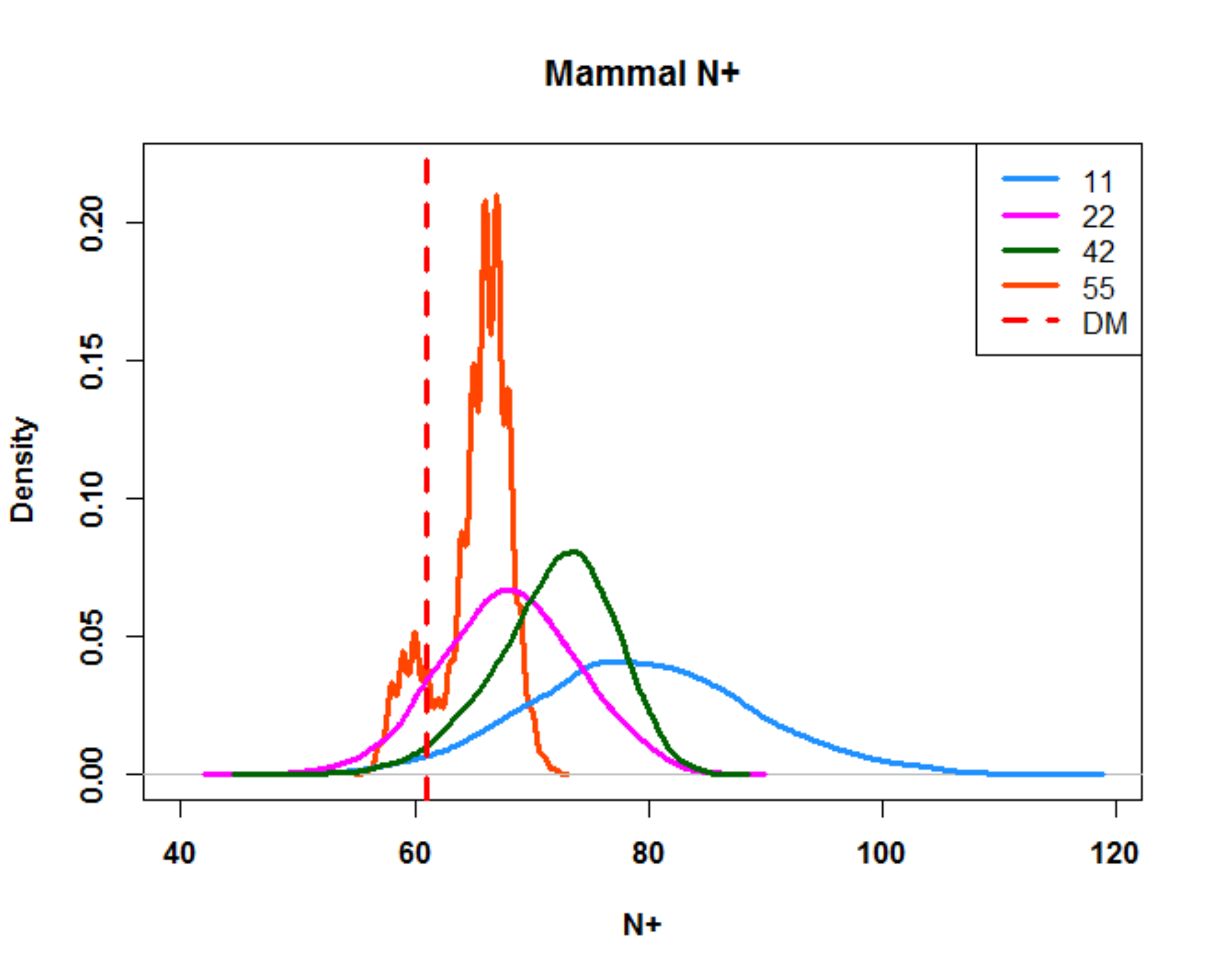}}
\subfloat[]{\includegraphics[trim=0cm 0.3cm 0cm 1.1cm, clip, width=3.5in]{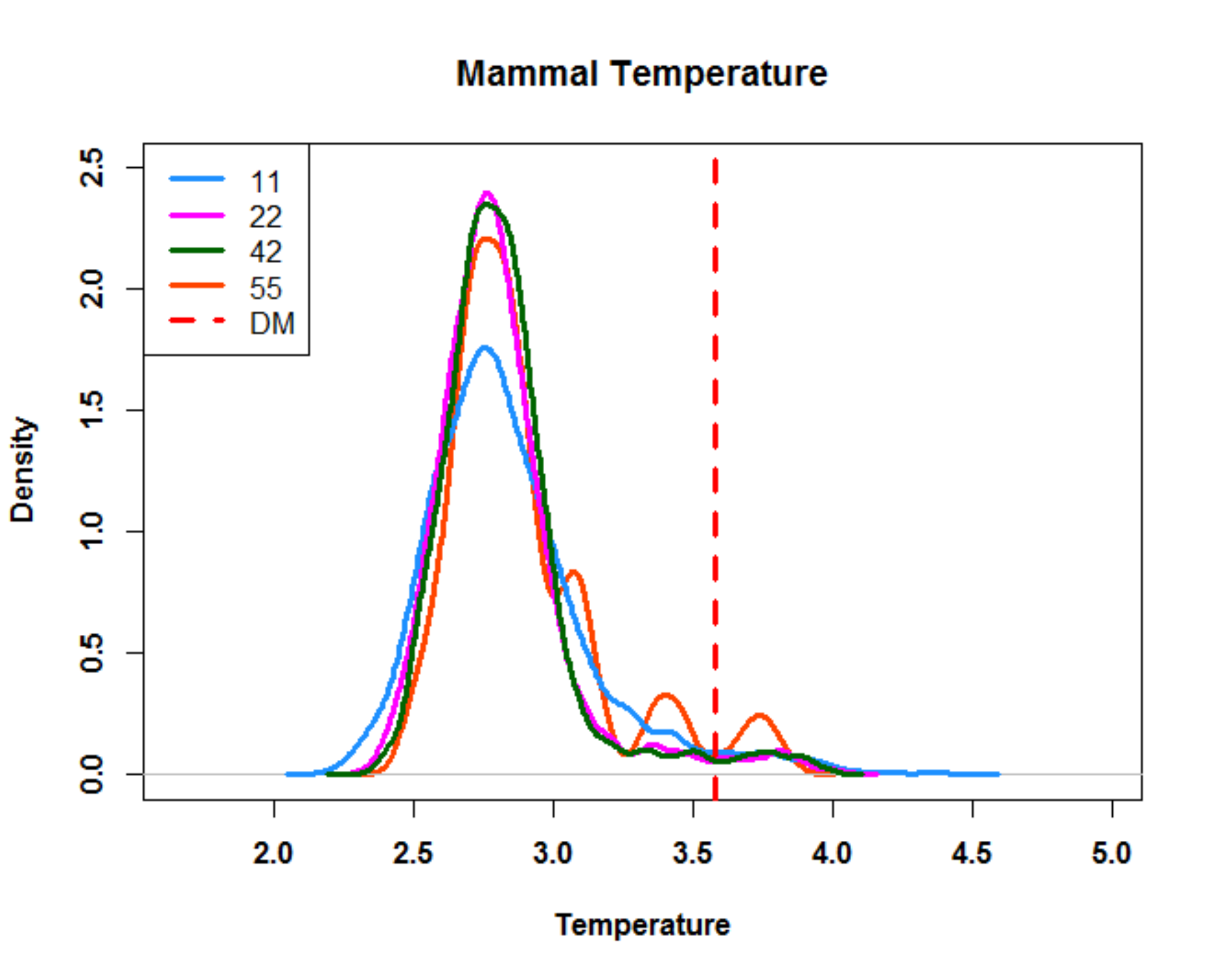}}\\
\subfloat[]{\includegraphics[trim=0cm 0.3cm 0cm 1.1cm, clip, width=3.5in]{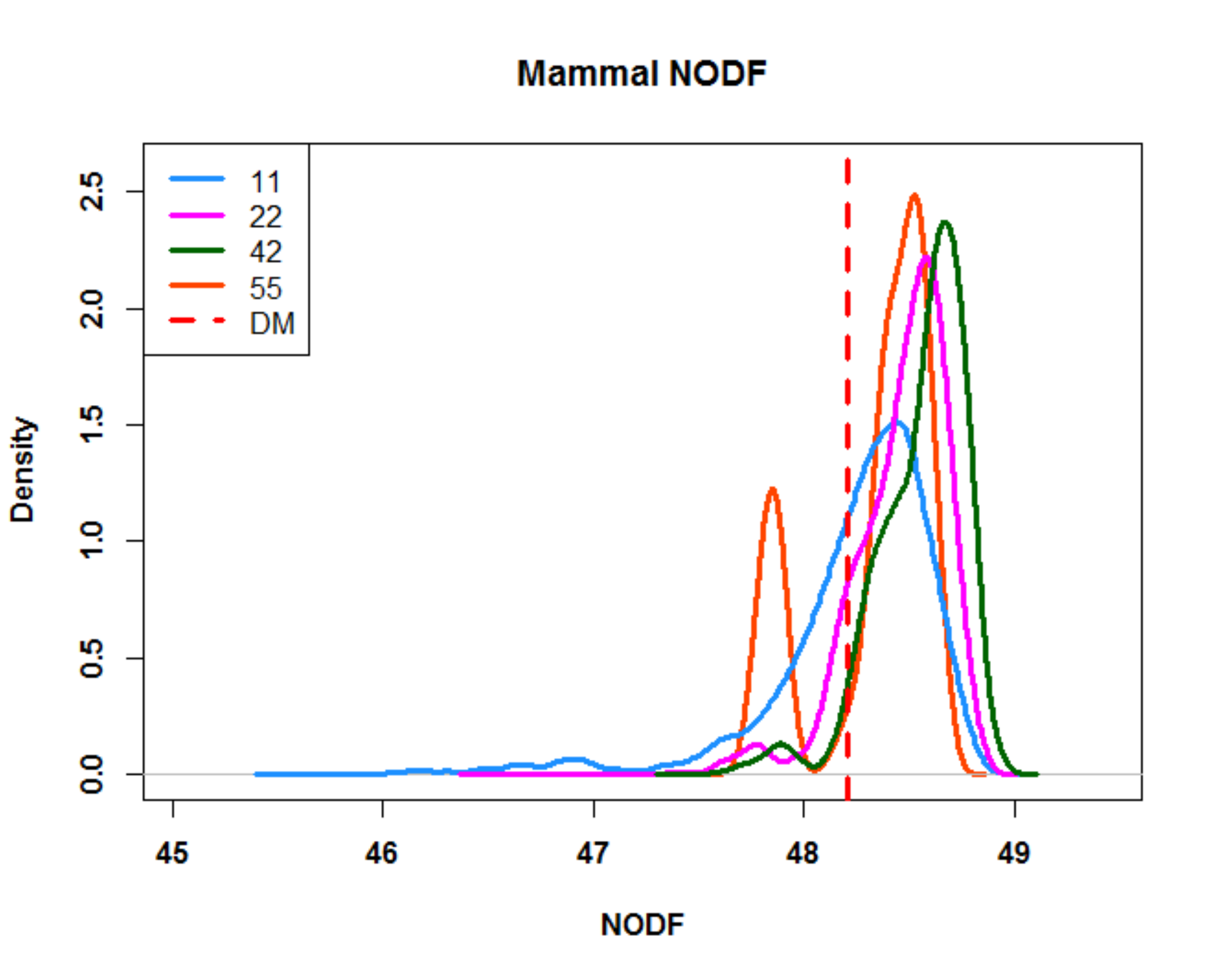}}
\subfloat[]{\includegraphics[trim=0cm 0.3cm 0cm 1.1cm, clip, width=3.5in]{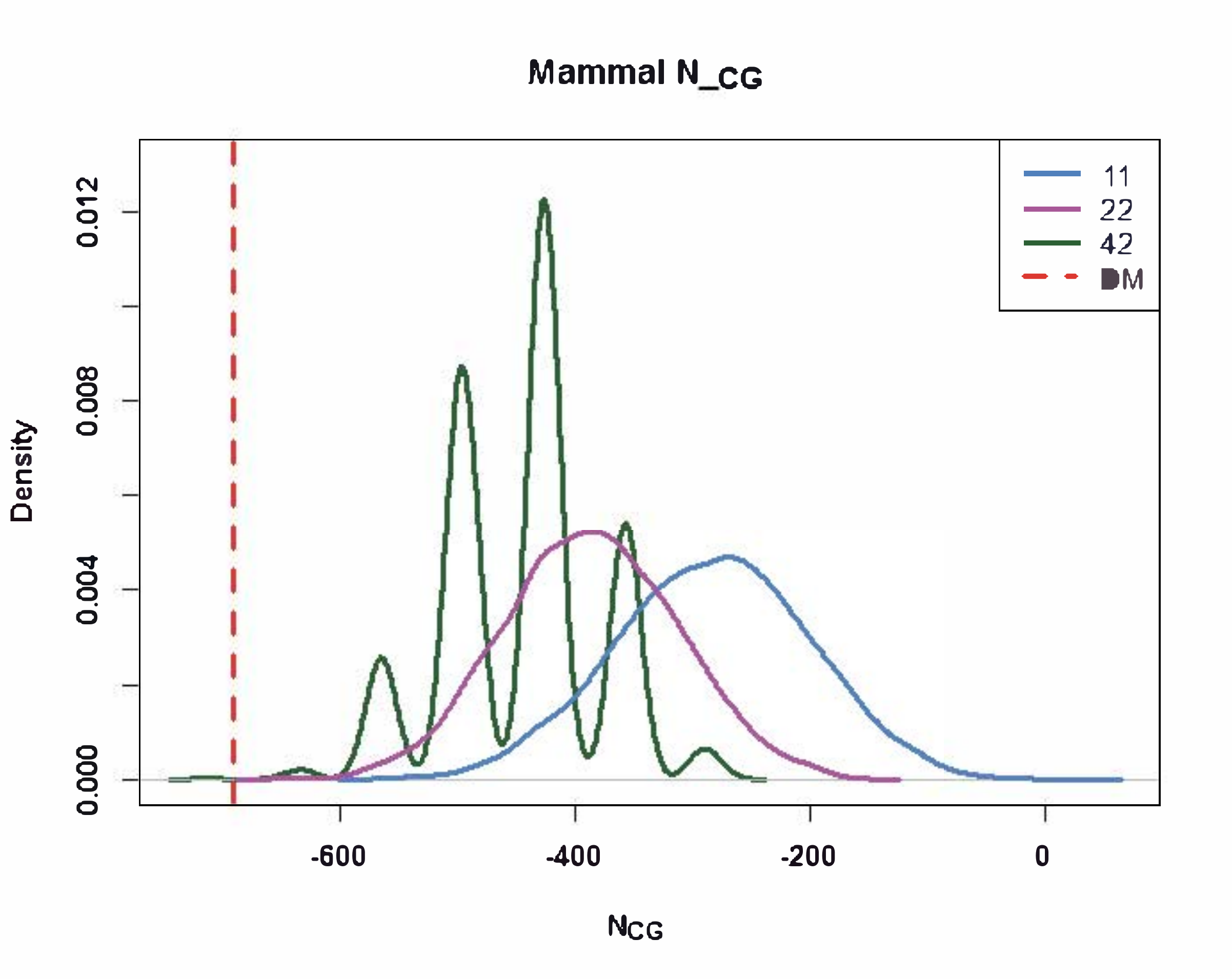}}

\caption{The distribution of different indexes which are sampled constrained by different scales of block structures. Each index and each scale of block structure are sampled 5,000 times using Miller’s algorithm. The dash line represents the indexes of Data Mechanics. }
\label{fig:Figure5}
\end{figure*}

So far there are at least three nestedness indexes have been proposed in literature. They are “$N+$ counts” (\cite{patterson1986nested}), T(temperature)-  (\cite{atmar1993measure}) and NODF (\cite{almeida2008consistent}) indexes.  The last two indexes are newly proposed and supposedly to improve the first index. However, as shown in the Fig. \ref{fig:Figure5}, these two supposedly improved versions are indeed “improper”. On the contrary, though it might not be effective, the originally proposed “$N+$ count” is not unreasonable.

Three existing indexes and the index NCG are computed upon four ensembles: $E_{5\times 5}$, $E_{4\times 2}$, $E_{2\times 2}$ and $E_{1\times 1}$, derived from the coupling geometry of Mammal data. Their corresponding distributions are presented in the following four panels. From the panel (a) for $"N+"$, though we see the distribution based on $E_{5\times 5}$ is somehow overlapping with the one based on $E_{1\times 1}$, while their two modes are evidently separated. And their relative positions are correct with the distribution based on $E_{5\times 5}$ as being more toward the nestedness and on the left of the one based $E_{1\times 1}$. In contrast, via panel (b) and (c) for T(temperature)- and NODF indexes, all their distributions are nearly completely overlapping with each other. Such complete overlapping phenomena strongly indicate that both indexes T(temperature)- and NODF are not effective statistics for testing nestedness given that the two ensembles $E_{5\times 5}$ and $E_{1\times 1}$ are very different in energy and size, so in pattern information. Finally, on the panel (d), the profile of distributions of $N_{CG}$ based on $E_{5\times 5}$, $E_{4\times 2}$, $E_{2\times 2}$  and $E_{1\times 1}$ is progressively shifting to the right as being away from nestedness. It is noted that the singleton based on ensemble $E_{5\times 5}$ is located at the extreme left tail of the one based on $E_{1\times 1}$. And it is understood that the 2nd order differences component in such an index is the key aspect that separate the coupling geometry and matrices in $E_{4\times 2}$, $E_{2\times 2}$ away from $E_{1\times 1}$.

Here we make a remark on the current state of knowledge: such an issue of how to systematically define an efficient testing statistic even on a binary bipartite network is still wide open at current state of knowledge. In fact it is expected because a bipartite network is indeed used to approximate a complex system state. This system state is very much nonparametric in nature in the sense of containing nonlinearity, dependence and heterogeneity.

\subsection{From coupling geometry to formulating the structural hypothesis  }
Finally we discuss two issues: 1) what are the rationales behind comparing the two distributions based on ensembles $E_{5\times 5}$ and $E_{1\times 1}$ in the Mammal data? 2) Why a hypothesis based on a binary bipartite network has to be formulated and tested in a conditional setting? These two issues are indeed two sides of the network data’s information contents.

First insight coming from the data’s matrix representation is that the row and column sums sequences might well be deterministic. That is, the two sequences involve with no randomness at all. Since it is what the system is at the moment the data being collected. Therefore conditioning on the two sequences is not only preferable, but necessary.

Via the conditioning thinking, the null and alternative hypotheses: Ho and Ha, are to be formulated as:

\textbf{Ho:} The mutualistic system of animal and plant interactions doesn't contain nestedness related patterns beyond the minimum patterns sustained by the row and column sums sequences.

\textbf{Ha:} The mutualistic system of animal and plant interactions does contain more nestedness related patterns in its minimum energy macrostate than the minimum patterns sustained by the row and column sums sequences.

Under Ho for the Mammal data, the null distribution with respect to any testing statistic is exactly the corresponding distribution derived from the ensemble based on $E_{1\times 1}$. However, under Ha, typically there exists one or many lowest energy macrostate embedded within an observed binary bipartite network data. For instance, we can find more than 300 matrices with lowest energy for the Mammal data. However explicitly finding minimum energy macrostates can be impractical due to huge computational loads. This is usually the case for large data networks. Hence a computable coupling geometry is the pragmatic candidate, on one hand.

On the other hand, since the hypothetic geometry specified on the alternative has to contain the minimum energy macrosate, so that an efficient testing statistic has a function of the coupling geometry as a minimum sufficient statistics. That is why the testing statistics is defined based on the finest scale block patterns. Such testing statistics are nearly identical with the ones defined on the real minimum energy microstates. Therefore the distribution under Ha pertaining to any efficient testing statistics has to be singleton based on the finest scale blocks, such as $E_{5\times 5}$ in Mammal data.

Nonetheless, it is critically important to emphasize here that any non-efficient statistics will have many “observed” values under the alternative hypothesis, such as  index on the Mammal data on ensemble $E_{5\times 5}$. So there would be a distribution of “p-values”. That is, the report of one single p-value is not valid in such a hypothesis testing setting. We indeed need to report such a distribution of P-values.

\section{Discussion}
We computationally extract a coupling geometry, consisting of deterministic and stochastic structures, embedded within an observed binary bipartite network as its information contents. From physical perspective, it is a minimum energy macrostate, and at the same time, from statistic perspective, it is the minimum sufficient statistic. Therefore any microstates as mimicries of the observed data network have to conform to this coupling geometry, while any potentially efficient testing statistics have to be a function of it. Further the pertinent geometric structure of nestedness has to be the least construct containing such a coupling geometry. These are fundamental facts underlying any coherent data analysis on binary bipartite networks. Significant implications include that the formulations of hypotheses are needed to be based on the minimum energy macrostate. And any potential nestedness index has to be in a form based on block patterns found on the finest scale.

The computable coupling geometry also facilitates various ensembles of matrix-mimicking according to its multiscale block patterns. Because of the profile of ensembles bearing with monotonically less geometric structures, any reasonably effective nestedness index will give rise to gradually separating index-based distributions: from the finest scale to the coarsest scale. That is to say that an index is not effective if it misses such a gradually separating pattern. When an index, such as $N_{CG}$ proposed here, is defined based on the finest scale blocks, it gives rise to singleton on the finest scale ensemble. Otherwise there would be a distribution of P-values.

On the front of generating random matrix subject to the two sequences of row and column sums, experiences from our computer experiments reveal that the commonly used $2\times 2$ checkerboard swapping and its variants need “big” perturbations in order to achieve “more uniform” sampling.  A perturbation-aided sampling scheme, based on a coupling geometry and Miller and Harrison’s algorithm (\cite{miller2013exact}) can generate and sample large random matrices up to thousand-by-thousand in size.

As a final remark, a coupling geometry computed from a binary bipartite network data can further afford an approach to compare the marginal tree structure on one axis with its corresponding phylogenetic tree as a new way of evaluating phylogenetic effects.  Such a comparison of two trees structures can be performed via a technique called, partial coupling geometry, which is developed in the spirit of mutual information in information theory.




\end{document}